\documentclass[reprint,aps,prb,amsmath,amssymb,showpacs,superscriptaddress]{revtex4-1}
\usepackage[colorlinks=true,linkcolor=blue]{hyperref}
\usepackage{graphicx}
\usepackage{dcolumn}
\usepackage{bm}
\usepackage{color}
\usepackage{upgreek}
\begin{document}
\title{Mechanism for femtosecond laser-induced periodic subwavelength structures on solid surface: surface two-plasmon resonance}
\author{Hai-Ying Song}\email{hysong@bjut.edu.cn} \affiliation{Strong-field and Ultrafast Photonics Lab, Institute of Laser Engineering, Beijing University of Technology, Beijing 100124, China}
\author{Shi-Bing Liu} \email{sbliu@bjut.edu.cn} \affiliation{Strong-field and Ultrafast Photonics Lab, Institute of Laser Engineering, Beijing University of Technology, Beijing 100124, China}
\author{H. Y. Liu}\affiliation{Max-Planck Institute for the Structure and Dynamics of Matter, Hamburg, Germany}
\author{Yang Wang}\affiliation{Strong-field and Ultrafast Photonics Lab, Institute of Laser Engineering, Beijing University of Technology, Beijing 100124, China}
\author{Tao Chen}\affiliation{Strong-field and Ultrafast Photonics Lab, Institute of Laser Engineering, Beijing University of Technology, Beijing 100124, China}
\author{Xiang-Ming Dong}\affiliation{Strong-field and Ultrafast Photonics Lab, Institute of Laser Engineering, Beijing University of Technology, Beijing 100124, China}
\date{\today}
\begin{abstract}
We present that surface two-plasmon resonance (STPR) in electron plasma sheet produced by femtosecond laser irradiating metal surface is the self-formation mechanism of periodic subwavelength ripple structures. Peaks of overdense electrons formed by resonant two-plasmon wave pull bound ions out of the metal surface and thus the wave pattern of STPR is ``carved" on the surface by Coulomb ablation (removal) resulting from the strong electrostatic field induced by charge separation. To confirm the STPR model, we have performed analogical carving experiments by two laser beams with perpendicular polarizations. The results explicitly show that two wave patterns of STPR are independently carved on the exposure area of target surface. The time-scale of ablation dynamics and the electron temperature in ultrafast interaction are also verified by time-resolved spectroscopy experiment and numerical simulation, respectively. The present model can self-consistently explain the formation of subwavelength ripple structures even with spatial periods shorter than half of the laser wavelength, shedding light on the understanding of ultrafast laser-solid interaction.
\end{abstract}
\maketitle
\section{Introduction}
Periodic ripple structures engraved inside/on any solid surfaces by means of femtosecond laser (fs-laser) pulses have been attracting a wide spectrum of scientists owing to its extensive applications in nano-optics, fabrication of electronic device, material physics, chemical surface etching etc. In morphology, when solid surface is irradiated with fs-laser pulse in air/vacuum, ripples are self-formed with an orientation perpendicular to the polarization of the employed laser and with a structuring spatial period (SSP) $\Lambda\!<\!\lambda_0$, or even $\Lambda\!<\!\lambda_0/2$ \cite{Sipe,Borowiec,Bonse}. This periodic structure in subwavelength scale unfolds renewed interest that it not merely allows the structured or characterized periods on the nanometre scale, but also has enabled us to control the structured patterns to reveal new aspects of their underlying peculiar functions. Thus it is of significant importance to exactly understand the formation process of fs-laser-induced periodic subwavelength structures on solid surface.

As it is well-known, SSP is experimentally proved to be determined by laser parameters (such as energy fluence $F_L$, wavelength $\lambda_0$, pulse duration $\tau_0$, and environment of processing) and material properties. The formation of ripples with SSPs close to or somewhat smaller than $\lambda_0$ is universally understood as an interference between the incident fs-laser beam and a surface electromagnetic wave (SEW) generated at rough surface (so-called surface plasmon polariton, SPP) \cite{Sipe,Borowiec,Bonse,AVorobyev,Miyaji,MHuang,JBonse,Garrelie}. In the case of normal incidence, SSP is usually evaluated by $\Lambda=\lambda_0/n$ \cite{Sipe,Emmony,Guosheng,Young}, where $n$ represents the refractive index of the dielectric material. This mode agrees with experimental results with $\Lambda\sim\lambda_0$ in a narrow fluence regime, however, it fails to interpret the formation of ripples with $\Lambda<\lambda_0/2$. Other possible mechanisms have been proposed and discussed by taking into account few more inferences, including self-organization \cite{Reif,Henyk}, second-harmonic generation \cite{Bonse, JBonse,Borowiec}, and interaction between laser pulses and SPPs \cite{Vorobyev4, Bonchbruevich,YHuang}. The origin of periodic surface ripple structures with subwavelength scale and its orientation remain a matter of great debate \cite{Bonse2012}, due to extremely intricate interaction of ultrafast laser with solids in the self-formation process, and hence, further investigations are indispensable for this subject.

In Table I, we summarize some experimental results have been reported before, including the ratios of SSP to laser wavelength, along with energy fluences and pulse durations used for different target materials. We also calculate the corresponding laser intensities $I_0$ by the relation $I_0=F_L/\tau_0$. One can see that the intensity values aggregate in the range of $5.6\times 10^{11}\!\sim\!8.75\times 10^{14}$ W/cm$^2$ which covers the regime of some parametric instabilities and reaches the first ionization threshold for majority of solid materials. According to Keldysh theory \cite{Keldysh}, the avalanche ionization (by electron impact) and multiphoton ionization will dominate at such intensity values in laser-solid interaction, which consequentially leads to an electron plasma sheet (EPS) formed close to the target surface with underdense density and lower electron temperature determined completely by material properties and laser parameters.

In this paper, we investigate the underlying mechanism for the self-formation of periodic subwavelength surface structures on solid surface induced by fs-laser pulses. We propose that surface two-plasmon resonance (STPR) driven by fs-laser in EPS, gives birth to periodic subwavelength-ripple structures. Specifically, under the instantaneous strong electrostatic field due to the separation of positive and negative charges (SPNC), peaks of overdense electrons formed by resonant plasmon wave pull bound ions out solid surface (i.e., ablating process). Thus the plasmon wave pattern on subwavelength scale is ``carved" on the surface of target material. On the basis of the STPR model, we have derived analytical formulas which agree with the present and previous observations in structuring experiments.
\begin{table}
\caption{Partial data for periodic surface structures reported on different solid targets: ratios of SSP to wavelength ($\Lambda/\lambda_0$), energy fluences [J/cm$^2$], pulse durations [fs], and corresponding laser intensity [W/cm$^2$] values.}
\begin{ruledtabular}
\begin{tabular}{cccccc}
 targets & $\Lambda/\lambda_0$ & $F_L$ & $\tau_0$ & Ref. & $I_0$[$\times10^{12}$] \\*[0.06cm]
 \hline
            &  0.88        &      0.25    & 100  &  [\onlinecite{Tsukamoto}]   &  2.5         \\
            &  0.66        & 0.067-0.084  & 65   &  [\onlinecite{Vorobyev}]    &  1.03-1.29   \\
    Ti      &  0.625-0.88  & 0.09-0.45    & 160  &  [\onlinecite{Okamuro}]     &  0.56-2.8    \\
            &  0.645-0.847 &     0.13     &  30  &  [\onlinecite{Bonse2012}]   &  4.3         \\
            &  0.68        &     0.5      & 500  &  [\onlinecite{Oliveira}]    &  1.0         \\
 \hline
    Cu      &  0.5-0.847   &  0.15-2.0    & 70, 100 &  [\onlinecite{Sakabe}]   &  2.14-2.8    \\
            &  0.338       &  0.04-0.1    & 70, 100 &  [\onlinecite{Sakabe}]   &  0.57-1.43   \\
 \hline
    Al      &  0.675       &  0.05        & 65      &  [\onlinecite{Vorobyev2}]  & 0.77       \\
 \hline
    Ni      &  0.75        &  0.12        & 50      &  [\onlinecite{Zuhlke}]     & 2.4        \\
 \hline
    W       &  0.775-0.88  &  0.2-1.1     & 160     &  [\onlinecite{Okamuro}]    & 1.25-6.875 \\
            &  0.5-0.75    &  2.5-7.0     & 33      &  [\onlinecite{Zhao}]       & 7.5-2.12   \\
 \hline
    Mo      &  0.775-0.88  &  0.2-1.1     & 160     &  [\onlinecite{Okamuro}]    & 1.25-6.875 \\
 \hline
    Au      &  0.72        &  0.16        &  65     &  [\onlinecite{Vorobyev3}]  & 2.46       \\
 \hline
    Pt      &  0.75-0.88   &  0.18-0.44   &  160    &  [\onlinecite{Okamuro}]    & 1.125-2.75 \\
            &  0.69-0.775  &  0.16        &  65     &  [\onlinecite{Vorobyev3}]  & 2.46       \\
 \hline
   SS30L    &  0.81        &  0.16        &  90     &  [\onlinecite{Yao}]        & 1.78       \\
 \hline
            &  0.625       &  0.08-0.2    &  130    &  [\onlinecite{Wu}]         & 0.615-1.53 \\
   AISI316L &  0.69        &  2.04        &  150    &  [\onlinecite{Bizi-Bandoki}] & 13.6     \\
            &  0.826       &  0.2-2.0     &  50     &  [\onlinecite{Zuhlke}]     & 4.0-40.0   \\
\end{tabular}
\end{ruledtabular}
\end{table}

\section{surface two-plasmon resonance}
When the surface of a solid target is normally exposed to a moderate intensity and linearly polarized fs-laser, the electromagnetic field of incident light that penetrates into the target can be treated as a solution of Maxwell equations coupled to the material equations, and the interaction of laser-material falls into the scope of the well-known skin-effect \cite{Rozmus,Luther}. The electrons in the skin layer $d_s\!=c/\omega_0\kappa$ are heated and ionized by the laser field of leading edge of the incident pulse, where $\kappa$ is the imaginary part of the refractive index in the Drude approximation and $\omega_0$ is the laser frequency. Almost simultaneously, part of the ionized electrons would brim over the surface and form an EPS with subcritical density near the target-surface with density scale-length $L_z\!=|n_e/(dn_e/dz)|$, where $n_e$ is the number density of electrons in EPS and $z$-axis is the propagating direction. In the subsequent interaction driven by the rest pulse, the oscillation of free electrons under the laser electric field $\textbf{E}_0$ generates two electron plasma waves (Langmuir wave) in opposite directions due to the inversion of the laser electric field, which is shown in Fig.1(a). Let the target-surface locate in $x$-$y$ plan $\textbf{r}_{\perp}$, then the density fluctuation of electrons associated with Langmuir wave (L-wave) due to the oscillation driven by the laser field can be concisely written as
\begin{align}
n'_e(\textbf{r}_{\perp},z,\,t)&=n_e(\textbf{r}_{\perp}\!+\textbf{r}_{os},z,\,t)
-n_e(\textbf{r}_{\perp},z,\,t) \notag\\
&\simeq\,\textbf{r}_{os}(\textbf{r}_{\perp},t)\cdot\nabla n_e(\textbf{r}_{\perp},z,\,t)\;,
\end{align}
where $\nabla\!=\partial/\partial\textbf{r}$ is a gradient operator and $\textbf{r}=(\textbf{r}_{\bot},z)$, $\textbf{r}_{os}\!=e\textbf{E}_0(\textbf{r}_{\bot},t)/(m_e\omega^2_0)$ is the spatial amplitude of the electron oscillation in the laser electric field $\textbf{E}_0(\textbf{r}_{\bot},t)$. Clearly from Eq.(1), the maximum fluctuation of the electron density (amplitude of L-wave) achieves at its gradient direction (wave vector $\textbf{k}$), i.e. parallel to laser electric field $\textbf{E}_0$.
\begin{figure}[!htb]
\includegraphics[height=7.6cm ,keepaspectratio=true,angle=0]{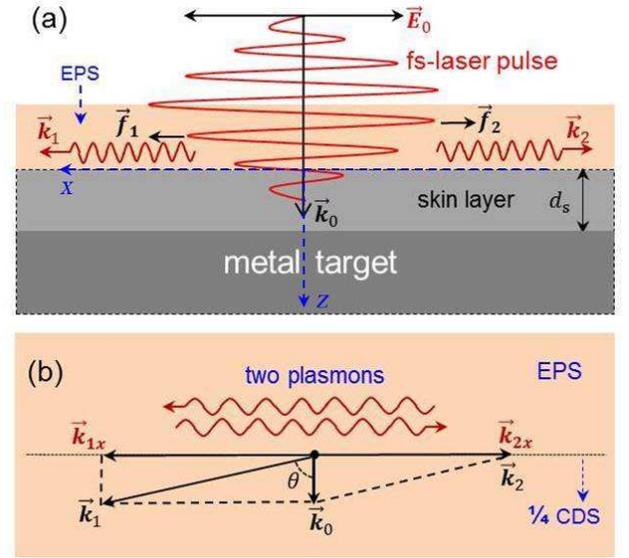}
\caption{\label{FIG1}(Color online) Schematic process of STPR driven by fs-laser pulse. (a)\,Generation of two daughter plasmon waves in EPS. (b)\,Wave vectors of the two plasmon resonances with matching conditions: $k_{1x}\!=\!-k_{2x}$ and $k_{1y}\!=\!k_0$. CDS denotes critical density surface.}
\end{figure}
Generally, when three waves encounter in the inhomogeneous plasma arious parametric instabilities may be driven \cite{Rosenbluth} as long as the phase relation of the three waves is matched, i.e.
\begin{align}
\omega_0=\omega_1+\omega_2\;\;\;    \text{and}   \;\;\; \textbf{k}_0=\textbf{k}_1+\textbf{k}_2 \,,
\end{align}
where $(\omega_0, \textbf{k}_0)$ is the pump light wave while $(\omega_{1,2}, \textbf{k}_{1,2})$ may be scattered light waves or plasma waves, or one scattered light wave plus one plasma wave. However, a direct instability process preferentially driven by laser electric field is that: a laser light wave resonantly decays into two L-waves, i.e., a photon\,$\rightarrow$ a plasmon + a plasmon. Since $\omega_1$ and $\omega_2$ are approximately $\omega_{pe}$, the scattered daughter L-wave has $\omega_1\!+\omega_2 \simeq 2\omega_{pe}\!=\omega_0$ ($\omega_{pe}\!\simeq\omega_0/2$\,), suggesting that a resonant two-plasmon decay process occurs at the neighborhood of the quarter-critical density surface ($n_c/4$, where $n_c\!=m_e/4\pi e^2\omega_0^2\simeq3\times10^{-10}\omega_0^2$ is the critical density). The corresponding dispersion relations for the three waves are then
\begin{align}
\omega_0^2=\omega_{pe}^2+c^2k_0^2 \;\;\;\;\;\;\;\;\; \text{for incident light wave} \\
\omega_{1,2}^2=\omega_{pe}^2+3v_e^2k_{1,2}^2 \;\;\;\; \text{for plasmon waves},
\end{align}
respectively, where $\omega_{pe}\!=\sqrt{4\pi e^2n_e/m_e}$ is the frequency of electron plasma and $v_e$ is the thermal velocity of electrons.

Assuming that a small perturbation of electron density gradient near the $n_c/4$ surface leads
one plasmon wave vector (such as $\textbf{k}_1$) to deviate from the direction of laser electric field or target-planar $\textbf{r}_{\perp}$ [\,see Fig.1(b)], then the change of $\textbf{k}_1$ would give rise to a small frequency shift, $(\,\omega_0/2-\omega_{pe}\,)=\pm\,\delta'$ and $|\delta'|\ll\omega_0/2$. In this case by using matching and dispersion relations, we gain
\begin{align}
\omega_0^2&=(\omega_1+\omega_2)^2\!=\omega_0^2/2+3v^2_e(k_1^2+k_2^2)-2\omega_0\delta'     \notag\\
&+\frac{1}{2}\sqrt{\omega_0^2\!+\!12v^2_ek_1^2\!-\!4\omega_0\delta'\,}\cdot\sqrt{\omega_0^2
\!+\!12v^2_ek_2^2\!-\!4\omega_0\delta'}  \;.
\end{align}
Expanding Eq.(5) and after daedal mathematical operation, the frequency difference of these two plasmon waves is obtained as
\begin{align}
\omega_1-\omega_2=3\omega_0\frac{v^2_e}{c^2}\frac{(\,k_1^2-k_2^2\,)}{k^2_0}
\end{align}
and $k_{1x}\!=\!-k_{2x}\!=-k_2$ [see Fig.1(b)]. One can find that, a small difference of $|\textbf{k}_1|$ and $|\textbf{k}_2|$ would result in a small frequency shift $(\,\omega_0/2-\omega_{1,2})=\pm\,\delta$ where ``+" means up-shift of the frequency and ``$-$" means down-shift of the frequency and $|\delta|\ll \omega_0/2$. Considering a small angle deviation of $\textbf{k}_1$ from target surface (i.e., $\theta \approx\pi/2$) due to the density fluctuation as shown in Fig.1(b), then the dispersion relation of plasmon $(\omega_1,\,\textbf{k}_1)$ becomes
\begin{align}
(\omega_0-\omega_1)^2=\omega_{pe}^2+3v_e^2\,(\textbf{k}_0-\textbf{k}_1)^2.
\end{align}
For convenience in writing, we omit the subscript ``1" of $\omega_1$ and $k_1$ hereinafter and thus have $(\omega_0-\omega)^2\!\simeq\omega_0^2/4+\omega_0\delta$ and $3v_e^2(\textbf{k}_0-\textbf{k})^2 \simeq\omega_0^2/4-\omega_{pe}^2\!+\omega_0\delta$, in terms of which the wavenumber of surface plasmon can be derived as
\begin{align}
k=\pm\,k_0\,\sqrt{1+\frac{4}{9}\frac{c^2}{v_e^2}\frac{\delta}{\omega_0}}\;\;,
\end{align}
where the sign $\pm$ means the daughter plasmon waves propagate along opposite directions. Herein one can see that the density perturbation is the essential condition for the three-wave coupling, as such the three waves can seek the phase matching state by self-organization in the coupling process and finally achieve phase-locked three-wave resonance. Once the phase-locked resonance is created, even if the incident laser pulse has ended, the wave model of STPR would keep oscillating with damping-amplitude way for a period of time just as a pendulum motion suddenly without external force. However in practice, the small perturbation of electron density or frequency is inevitable due to the unpredictable changes of the parameter conditions in the interaction of laser-targets, which undoubtedly results in nonzero density and frequency detunings ($\delta',\delta\!\neq 0$) in the three-wave coupling. Consequently, this STPR is easily to be driven only if the intensity of pumping laser light reaches threshold condition of the parametric instability growth. In view of this, considering the case of $\omega>\omega_{pe}$ (\,i.e., $\delta>0$ situation), we have
\begin{align}
\delta=\omega-\frac{\omega_0}{2}=\!\left[\sqrt{4(1+3k^2\lambda_D^2)/5\,}-1\right]\!\frac{\omega_0}{2}
=\mu\frac{\omega_0}{2} \;,
\end{align}
where $\lambda_D\!=v_{e}/\omega_{pe}$ is Debye length and characterizes the spatial characteristic scale. It has been demonstrated that the Langmuir decay instability associated with multiple L-waves generation is driven in the wave-wave nonlinear regime of $k\lambda_D<0.29$\,\cite{Kline}. Congruously, the change of kinetic parameter is caught in a very narrow range of $0.2886<k\lambda_D<0.29$ for STPR and thus we take the frequency shift factor $\mu=\frac{1}{2}\times10^{-4}$ for convenience. Consequently, according to Eqs.(8) and (9) the wavelength of STPR (using $\lambda/\lambda_0=k_0/k$) writes
\begin{align}
\lambda=\frac{\lambda_0}{\;\sqrt{1+1.1\times10^{-5}\,m_ec^2/T_e}\;\,}\,,
\end{align}
where $T_e$ is the temperature of electrons in EPS and $v_e\!=\!\sqrt{T_e/m_e}$. In fs-laser-produced plasmas, the electron temperature is mainly proportional to the applied laser intensity (or fluence) and is dependent on the absorption mechanism in the interaction.

\section{estimate of electron temperature}
Through Eq.(10), the wavelength of STPR is determined by not only the laser wavelength $\lambda_0$ but also the electron temperature $T_e$ which depends on the parameters of light and target material. For the interaction of laser with solid targets, the pulse duration of sub-picosecond appears to be shorter than all characteristic relaxations, such as energy transfer from electrons to ions and electron heat conduction. The major process for laser-target coupling is the electron heating by the laser field, and the interaction falls in the scope of the skin effect. In the meanwhile, the portion of ionized free electrons that overbrim the target surface oscillate in the laser electric field and simultaneously are damped in the electrostatic field formed by target-ions (random inelastic scattering between electron-ion). Therefore these underdense electrons have to be randomly heated instantaneously before the Coulomb ablation occurs. A significant coupling mechanism for laser-plasma interaction in underdense region ($n_e\!<n_{cr}$) is inverse Bremsstrahlung (IB) or collisional absorption \cite{JMDawson,TWJohnston,SBLiu}, leading to a conversion from laser into random thermal energy via the absorption of electronic kinetic energy. Thus the temperature of electrons in EPS, in general, is higher than that in the skin layer during the interaction of laser-targets. Whereas for a moderate intensity of incident fs-laser, the scale length of EPS is much smaller than the laser wavelength in such an ultrafast interaction process, and thereby we can substitute the surface electron temperature in skin layer for that in EPS approximatively. According to energy conservation, the equation for the electron temperature $T_e(z,\,t)$ due to the absorption in the skin layer presents \cite{WRozmus,BNChichkov}
\begin{align}
C_en_{e0}\frac{\partial T_e}{\partial t}=\alpha AI_0\exp(-\alpha z)\;,
\end{align}
where $\alpha=2/d_s$, $C_e$ is the heat capacity (per unit volume) of electrons, $A$ is the absorption coefficient and $A/d_s=2\omega_0/c$ \cite{WRozmus}, and $n_{e0}$ is the electron density in the skin layer for metal materials and can be substituted by atom density $n_a$ for nonmetal materials. Due to the fact that the skin layer is usually much thinner than the heated plasma region, we treat the laser energy absorption in the skin layer as a surface effect. Furthermore, the electron heat capacity increases with the increasing of electron temperature from a low temperature degenerate state $C_e\!=\frac{1}{2}\pi^2T_e/\varepsilon_{F}$ up to a maximum value of $C_e\sim 3/2$ for a conventional ideal gas \cite{CKittel}, where $\varepsilon_{F}$ is the Fermi energy of electrons.
\begin{figure}[!htb]
\includegraphics[height=6.5cm ,keepaspectratio=true,angle=0]{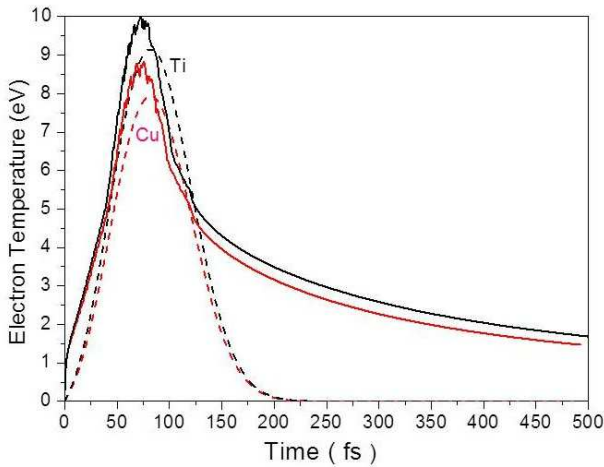}
\caption{\label{FIG2} (Color online) Transient electron temperatures calculated for Ti (black lines) and Cu (red lines) targets, using the present theory (dashed lines) and the Med103 code (solid lines).}
\end{figure}
Based on this physical argument we allow the same heat capacity as for an ideal gas because the ionization process has been fulfilled early in the laser pulse, and the electrons are in conditions close to that of an ideal gas. Consequently, the concise linear scaling relation for the electron temperature with respect to fluence at the surface of skin layer irradiated by laser pulses, according to Eq.(11), is obtained as
\begin{equation}
T_e(z=0,\,t=\tau_0)\simeq 10^{24}\times F_L/(\lambda_0n_{e0})\;,
\end{equation}
where the units of $T_e$ is in eV, $F_L$ in J/cm$^2$, $\lambda_0$ in $\mu$m, and $n_{e}$ in cm$^{-3}$, respectively.

In order to check the validity of the prediction, the hydrodynamic simulation for a Gaussian pulse incident to metal targets is performed by a one-dimensional Lagrangian code MEDUSA (Med103) \cite{Christiansen,ADjaoui}. The transient electron temperatures for Cu and Ti are obtained by the simulations with laser intensity of $10^{12}$\,W/cm$^2$ and wavelength of 0.8\,$\mu$m, as shown in Fig.2. One can see that the maximum temperatures obtained from the present prediction are lower than these from the hydrodynamic simulation and also emerge at later moments for both targets. As for these differences, a straightforward reason is that, as mentioned above, Eq.(11) is based on heat diffusion mechanism in materials, and the defined temperature is limited in the skin layer, while the Med103 code is based on IB absorption mechanism in which the electro-ion scattering effect is taken into account in the heating process.

\section{ablation by STPR wave mode}
Although the matching conditions in Eq.(2) allow a broad wave-number spectrum of plasmon waves, the wave mode with $\textbf{k}\|\textbf{E}_0$ would be preferentially developed, as shown in Eq.(1). We further examine the growth rate of two-plasmon decay process when three-wave-resonance takes place, i.e. \cite{Rosenbluth}
\begin{align}
\gamma=\frac{\textbf{k}\cdot\textbf{v}_{os}}{4}\left[\frac{(\textbf{k}-
\textbf{k}_0)^2-k^2}{k|\textbf{k}-\textbf{k}_0|}\right] \,,
\end{align}
where $\textbf{v}_{os}\!= e\textbf{E}_0/(m_e\omega_0)$ is the quiver velocity of electron in laser field and $\textbf{k}$ marks both $\textbf{k}_1$ and also $\textbf{k}_2$ for convenience. It indicates that, only when $\gamma>0$ is satisfied the two-plasmon decay process can develop resonantly and the amplitude of plasmon wave grows steadily. Obviously from Eq.(13) and phase matching relations, the condition of $\gamma>0$ requires $k<\frac{1}{2}\,k_0/\cos\theta$. Then the plasmon mode with wave vector approximatively paralleling to the target-planar (large $\theta$ angles) has a vector $k\gg k_0$, giving rise to a maximum growth rate $\gamma_m\!=k_0v_{os}/4\simeq 6.25\times 10^{-4}\sqrt{I_0}$, where $I_0$ is the amplitude of the laser intensity (in W/cm$^2$). It shows that a lower intensity threshold of pumping laser is allowed for driving such a parametric resonance of the surface two-plasmon near $n_c/4$. The growing amplitude of the plasmon wave means the increased charge density of the electronegative center in EPS, $\rho=-en'_e$, resulting in a charge separation as shown in Fig. 3. Consequently, a periodic interface electrostatic field (IEF) is effectively formed on the gradient of the electron density along the normal to the target surface (assuming one dimensional expansion), associating with the amplitude of plasmon wave in EPS. On the basis of such a charge separation, the electrostatic field intensity can be expressed as \cite{Gamaly}
\begin{align}
E_z(x)=-\frac{\varepsilon_{ek}}{e}\cdot\frac{\partial(\ln n_e)}{\partial z}\simeq \frac{F_L}{ed_sn_e\lambda_D}\,,
\end{align}
where $\varepsilon_{ek}\!=m_ev_{e0}^2/2\,(=\varepsilon_{abs}-\varepsilon_w)$ is the kinetic energy of the energetic electrons, $v_{e0}$ is the velocity of electrons rushing out the target surface during the ionization process, $\varepsilon_{abs}$ is the total energy absorbed by electrons from laser pulse, $\varepsilon_w$ is the work function, and $\lambda_D$ is the Debye length. Thus, once the excitation energy of this electrostatic field force at the peaks of electronegative centers to the target-ion, $eE_z(x_i)\lambda_D$, exceeds its binding energy in the lattice, the target-ions at $x_i$ would be pulled out the target surface by Coulomb ablation (also known as Coulomb explosion or phase explosion) as shown in the schematic diagram in Fig.3.
\begin{figure}[!htb]
\includegraphics[height=3.9cm ,keepaspectratio=true,angle=0]{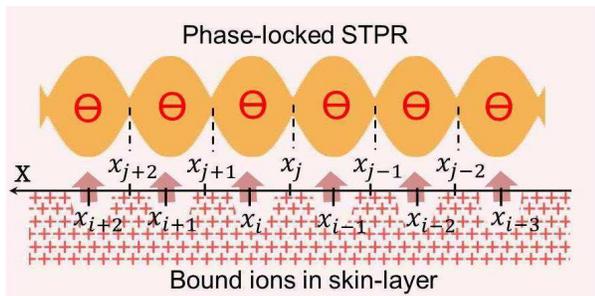}
\caption{\label{FIG3}(Color online) Schematic mechanism of phase-locked STPR wave producing periodic Coulomb ablation. ``\,$\ominus$" at coordinates $(...,\,x_{i-1},\,x_{i},\,x_{i+1},\,...)$ labels the electronegative centers formed by STPR due to the peaks of wave electrons, and ``\,+" denotes the bound ions in the skin-layer. Coulomb ablation occurs only at the peaks of overdense wave electrons due to the electrostatic field between electronegative centers and target ions.}
\end{figure}
Consequently, analogous to traditional carving arts by removal material, the wave pattern of STPR is ``carved" on the target surface by this Coulomb ablation mechanism. The SSP of the structured ripples is equivalent to the wavelength of STPR ($\Lambda=|x_{i+1}-x_i|=\lambda$\,) and the scaling equation, according to Eqs.(10) and (12), is obtained as
\begin{align}
\Lambda=\frac{\lambda_0}{\sqrt{1+5.62/T_e}\;\,}\,,
\end{align}
where the unit of $T_e$ is in eV. It shows that the SSP depends mainly on the laser wavelength and the properties of target materials (temperature dependence). Here one can see that the SSP is explicitly in the subwavelength regime unless an extremely high laser fluence or electron temperature is used. It needs to indicate that the scaling equation (15) is also appropriate for dielectric targets, for which the electron temperature $T_e$ should be replaced by dependence relation of laser-dielectric materials. Furthermore, in terms of equations (1) and (5) we also can easily conclude $\textbf{k}\parallel \textbf{E}_0$ which means that the orientation of ripples is perpendicular to the polarization of incident laser pulses.

In order to assess the time necessary to fulfill the Coulomb ablation (i.e., Coulomb explosion), we define the Debye length as a basic distance over which the target ion is considered to be dragged out of the lattice. Therefore, the energy condition for Coulomb ablation accords with $m_iv_{i0}^2/2\!=eE_z(x_i) \lambda_D\!-(\varepsilon_b+\varepsilon_w)\!>0$ and the change of ion momentum meets Newton's law $m_i\text{d}v_{i0}/\text{d}t\!=eE_z(x_i)$, where $\varepsilon_b$ is the binding energy of ions in the lattice and $v_{i0}$ is the initial velocity of ions out of the lattice. Consequently, the time necessary to accelerate and ablate an ion can be roughly evaluated by the relation of
\begin{align}
t_{acc}=\lambda_D/v_{i0}=\omega_{pe}^{-1}\frac{v_{e0}}{v_{i0}}\;.
\end{align}
As an example to conservatively estimate the acceleration time of an ion in copper target ($\varepsilon_w\!=4.65$\,eV and $\varepsilon_b=3.125$\,eV \cite{CKittel}) at wave peaks $x_{i\pm\cdots}$, we consider that the electron density at $x_{i\pm\cdots}$ (wave peaks of STPR mode) is ten times the critical density and the absorbed energy $\varepsilon_{abs}$ is approximatively replaced by electron Fermi temperature (\,$\approx$ 7.0\,eV\,), consequently, this time gives less than 40\,fs at $F_L\!=0.5$\,J/cm$^2$ and  $\lambda_0\!=0.8\,\mu$m ($d_s\!\approx 67$\,nm). In this case, it indicates that the Coulomb ablation is possibly achieved as long as the lifetime of STPR state remains over 40\,fs.

It needs to emphasize that the periodicity of surface structures with wave pattern will disappear (taking on crater shape) if the electrostatic field forces at STPR valleys $x_{j\pm\cdots}$ are so strong so that bound ions can be effectively dragged out of target surfaces. Therefore, the applied laser fluence in structuring experiments is necessarily to be limited below an upper ablation threshold, which will be discussed in another paper.

\section{experimental verification}
In order to confirm the present physical model we carried out experimental verification. Our experimental setup is equipped with a commercial fs-laser system (\,Ti\,:\,sapphire Micra 10 and regenerative amplifier Legend Elite-USP-HE, Coherent Corp.\,) with 60\,fs pulse duration, 800\,nm central wavelength, 3.5\,mJ/pulse, and repetitive rate of 1\,kHz. To implement time-resolved spectroscopy measurements from fs-laser structuring, we used an ICCD camera (PI Corp., USA) with a basic gate width (i.e.exposure time for one triggering) of $\tau_\Delta\!=2$\,ns.

\subsection{Lifetime of surface plasmon induced by fs-laser}
As mentioned above, the lifetime of the surface plasma state produced in the interaction of laser-targets should be long enough in order to realize the periodic Coulomb ablation by the phase-locked STPR. We know that laser-produced plasma (charged particles) states would emit abundant electromagnetic radiations continuously (continuous spectrum, CS) due to their random motions, and while the electrons begin to recombine with their parent ions the characteristic line spectra (CLS) radiate because of the state transition between energy levels in atoms. Therefore, the duration $\delta t_p$ from the emission of CS to the appearance of the first CLS characterizes a pure plasmon state. In other words, the total number of free electrons remains unchanged during the time from the end of laser pulses to the emission of the first CLS. Here we define $\delta t_p$ as the lifetime of surface plasmon, which is the basis for the existence of plasmon wave (as well as the STPR). In view of this feature, to evaluate the duration $\delta t_p$, we have performed a series of time-resolved spectroscopy measurements. The camera shutter is triggered by the trigger signal output from the fs-laser system, and the emitted spectra are guided into a spectrometer through a grating and detected by a ICCD camera.

In order to generate optimized periodic surface ripple structures, firstly, a linearly polarized fs-laser with a fluence of 0.4\,Jcm$^{-2}$/pulse is applied at a fixed position on Cu-metal surface. After successively irradiated by 100 pulses, the scanning electron microscope (SEM) image is shown in Fig.4(a).
\begin{figure}[!htb]
\includegraphics[height=3.9cm,keepaspectratio=true,angle=0]{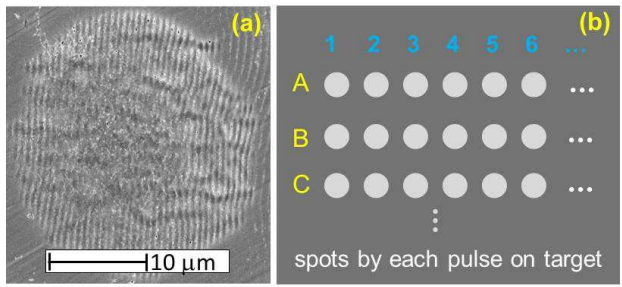}
\caption{\label{FIG4} (Color online) Schematic illustration of spot matrix for the measurement of time-resolved spectra. (a)\,SEM image taken on Cu target in fs-laser structuring experiment. (b)\,Spot matrix for time-resolved spectra on target shown in (a). Spots with different shutter/trigger time are grouped in A, B, C etc.}
\end{figure}
The time evolution of the emitted spectra can be resolved by varying the trigger time continuously. In addition, to minimize the inhomogeneous effect, different spots on the target surface are measured, forming a spot matrix, as shown in Fig.4(b). Specifically, each group has the same trigger time, and tens of spectra (noted as 1, 2, 3, 4, 5, 6,...) at different spots in the same group are integrated. For example, group ``A" contains spectra at trigger time $t_1$, while the second group ``B" is triggered at $t_2$, then a time interval $\Delta t\,(\,=t_2-t_1)$ can be obtained. Figure 5 shows the measured time-resolved spectra from Cu target with a time interval of 1\,ns. One can find that the first CLS (\,$5\text{s}'\,^4\text{D}_{7/2}\rightarrow 4\text{p}'\,^4 \text{F}^0_{9/2}$, 465.11\,nm\,) \cite{HCederquist} shows up at the 10th spectral curve. Therefore, in terms of time-resolved spectra we can conclude that the lifetime of surface plasmon (electron) approximates 10\,ns which is at least five orders of magnitude longer than the acceleration time of target-ions.

\subsection{Carving effect of STPR wave patterns}
In order to verify the carving effect resulted from STPR mechanism, we carried out structuring experiments using two beams, as illustrated in Fig.6(a). The sample surface (W-target) was only simply polished mechanically to avoid too high reflectivity, and the pulse profile was shaped properly. The average power of the incident laser can be altered from 1\,mW by using a variable density filter, and the pulse account is controlled by an synchronized electromechanical shutter (not shown in the picture).
\begin{figure}[!htb]
\includegraphics[height=5.6cm ,keepaspectratio=true,angle=0]{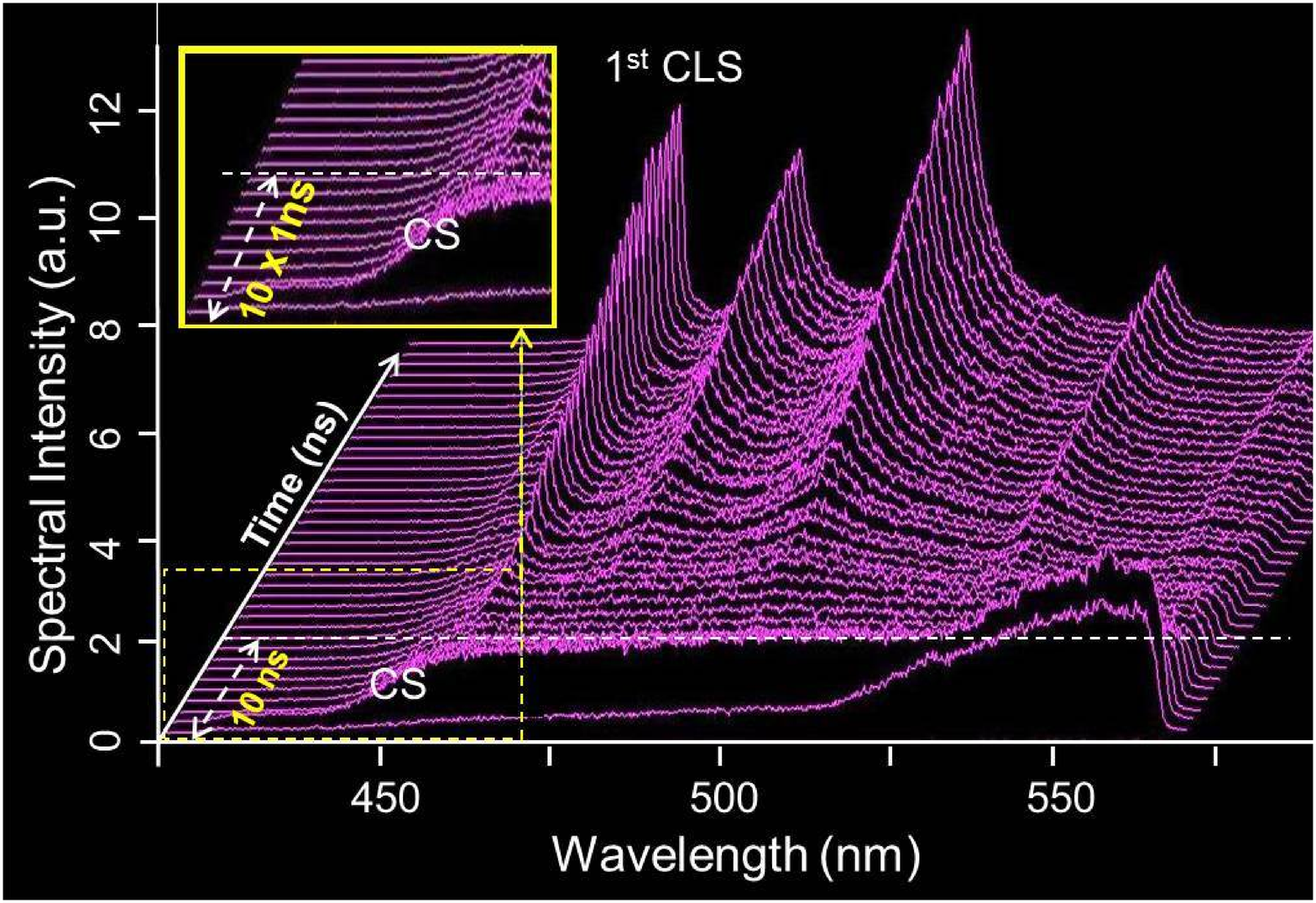}
\caption{\label{FIG5} (Color online) Time-resolved spectra measured on Cu target by irradiation with linearly polarized 60 fs laser pulses with a central wavelength of 800 nm and a fluence per pulse of 0.4\,Jcm$^{-2}$.}
\end{figure}
To show the carving effect of different wave patterns induced by STPR, the linearly polarized fs-laser beam is equally split in two beams by a beam splitter, denoted as p-$\textcircled{\small{1}}$ and p-$\textcircled{\small{2}}$, with electric field intensities $\textbf{E}_{01}$ and $\textbf{E}_{02}$. The delay time between these two beam is tuned by a motorized time-delay-line (TDL) and the polarization direction of p-$\textcircled{\small{2}}$ ($\textbf{E}_{02}$) can be altered by a half wave plate ($\lambda/2$).

In the two-beam structuring experiment the delay time $\Delta\tau$ between p-$\textcircled{\small{1}}$ and p-$\textcircled{\small{2}}$ was chosen at 100\,ns so as to have $\Delta\tau$ much larger than the relaxation time of electron heat diffusion. The polarization directions of two beams were selected as mutually perpendicular (\,$\textbf{E}_{01}\!\perp\!\textbf{E}_{02}$\,), and the whole structuring process was performed in air. Thus, the observed results positively reveal the effect of alternant ablating of two wave patterns by STPR with $\textbf{E}_{01} \perp\textbf{E}_{02}$ (layer-carving effect).
\begin{figure*}[!htb]
\includegraphics[width=1.9\columnwidth,angle=0]{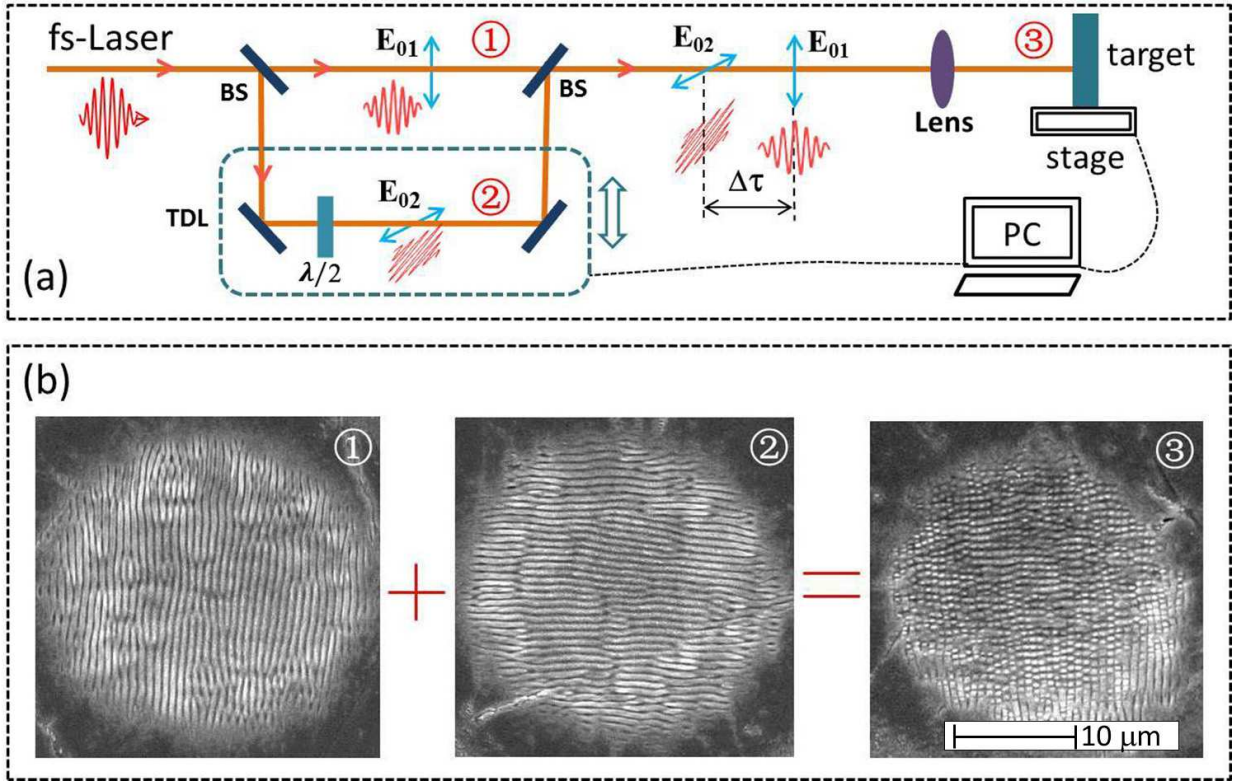}
\caption{\label{FIG6}(Color online) (a)\,Overview of the experimental setup for periodic ripple experiments using two beams with perpendicular light polarizations. BS is a non-polarizing 50/50\% beam splitter, $\lambda/2$ is a half wave plate, and TDL is a motorized time-delay-line. (b)\,Ripples (SEM) produced by pulses with horizontal ($\textcircled{\small{1}}$), vertical ($\textcircled{\small{2}}$), and dual ($\textcircled{\small{3}}$) polarizations. All pictures have a 10 $\mu$m scale.}
\end{figure*}

Figure 6\,(b) shows SME images under the irradiation of 0.6\,J/cm$^2$, indicating clear layer-carving effects: image $\textcircled{\small{1}}$ and image $\textcircled{\small{2}}$ are produced by beam p-$\textcircled{\small{1}}$ (horizontal polarization) and beam p-$\textcircled{\small{2}}$ (vertical polarization), respectively; while image $\textcircled{\small{3}}$ is a combination produced by p-$\textcircled{\small{1}}$ and p-$\textcircled{\small{2}}$. This demonstrates that the wave patterns for STPR driven by p-$\textcircled{\small{1}}$ and p-$\textcircled{\small{2}}$ beams are effectively carved by removal material on the target surface, the orientation of structured ripples is perpendicular to the polarization direction ($\textbf{E}_0$) of the driving laser, the SSP is located in subwavelength ($\sim 0.75\,\lambda_0$), and especially, the feature of STPR wave produced layer-carving is significantly highlighted here.

\subsection{Comparison of observed and predicted SSPs}
The self-formation of periodic surface structures by fs-laser pulses is determined by the underlying physical mechanism, with the structure morphology depending on material properties, laser parameters, and processing methods. However, even for the same materials and roughly identical laser fluences, SSP results from different literatures can deviate from each other, as one can see in Table I. These differences may arise from the estimation for the radius of laser focal spot, selection from their SEM images, processing environment, etc.
\begin{figure}[!htb]
\includegraphics[height=6.5cm,keepaspectratio=true,angle=0]{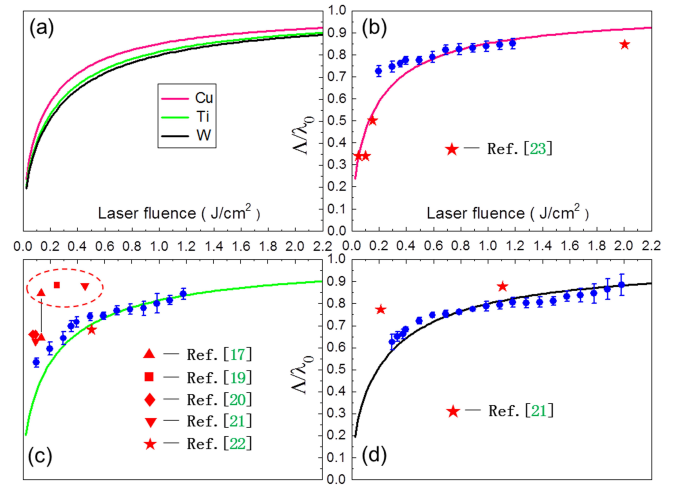}
\caption{\label{FIG7} (Color online) Ratios of ripple's SSP to laser wavelength ($\Lambda/\lambda_0$) vs laser fluence. (a)\,Theoretical redictions by Eq.(15) for Cu, Ti, and W targets. (b)-(d) Comparisons of results and predictions for Cu, Ti, and W targets. Blue dots are results from our experiments.}
\end{figure}
In order to compare experimental observations with theoretical predictions by Eq.(15), Figure 7 summarizes SSPs as a function of laser fluence for three material targets, Cu, Ti, and W, from both our measurements and previous results. In Fig.7(a), one can see small deviations for different materials in theory. For Ti target shown in
Fig.7(c), much larger deviations appear between the observations reported in Refs.[\onlinecite{Tsukamoto,Vorobyev,Okamuro}] and our prediction, however our calculation agrees very well with the results from our experiment and Ref.[\onlinecite{Oliveira}]. We should emphasize that, for Cu target shown in Fig. 7(b), the observed data are in good agreement with the prediction and especially, the present physical model can self-consistently explain the ripple structures with SSPs shorter than half of laser wavelength ($<\!\lambda_0/2$\,) at small laser fluences (\,$\leq 0.1$\,J/cm$^2$\,) \cite{Sakabe}.

\section{discussion and conclusion}
We propose here a new mechanism for surface structuring on solid surfaces by fs-laser pulses, and present an analogous surface carving notion arising from phase-locked STPR wave pattern. The interaction is characterized by laser intensity in the range of $10^{11}$ to $10^{14}$\,W/cm$^2$, with a pulse duration of tens of femtoseconds. We note the interaction time is much shorter than the plasma expansion time, the heat conduction time, the electron-ion energy transfer time, and especially the ion respond time in which the ion can be considered as an immobile electropositive background. Under such conditions the ionization of any target material is practically realized. The electron density perturbation (\,$\mu\neq 0$\,) facilitates the phase matching of three-wave resonance in EPS formed by hot electrons escaping from target surfaces. Therefore, L-wave ablation resulted from STPR is responsible for the self-formation of periodic subwavelength ripple structures. The notion of L-wave pattern carving (Coulomb ablation) occurs in the acceleration of target-ions at the L-wave peaks located in the periodic interface electrostatic field created between negative electricity centers and target ions. The wavelength of STPR is closely determined by electron temperature, as shown in Eq.(10). In the estimation of electron temperature in EPS, the loss of laser energy in ionization process is ignored and $T_e$ in surface plasmon is replaced by that in the skin layer of solids, which is valid only in the case of low laser intensities ($<10^{14}$\,W/cm$^2$\,).

In summary, we derived explicit analytical formulas for dispersion relations between phase-locked STPR, laser wavelength (wavenumber), SSP, and electron temperature in ultrafast interaction between fs-laser and solid surfaces. The relevant dynamics in the present model, including relaxation of surface plasmon, electron heating, and carving effect of L-wave pattern, were verified respectively by time-resolved spectroscopy, numerical simulations, and two-beam structuring experiments. The theoretical calculations agree with both our and previous experimental data, and especially can self-consistently explain SSP $\Lambda\!<\!\lambda_0/2$\ at low fluences. Our model can be readily extended to more materials including metals and dielectrics under fs-laser irradiation.

This work is supported by the National Natural Science Foundation of China (Grant No. 51275012) and NSAF of China (Grant No. U1530153), the Ministry of Science and Technology of China Major Project of Scientific Instruments and Equipment Development (Grant No. 2011YQ030112), and the Beijing Commission of Education (No. KZ201110005001).

\end{document}